\documentclass[a4paper,12pt]{article}
\usepackage{epsfig}
\usepackage{color}
\usepackage[numbers,sort&compress]{natbib}

\newlength{\dinwidth}
\newlength{\dinmargin}
\newcommand{\spur}[1]{\not\! #1 \,}

\begin{document}
\title{Decays
$D^+_{(s)}\to \pi(K)^{+}\ell^+\ell^-$ and $D^0\to \ell^+\ell^-$ in
the MSSM with and without R-parity}

\author{ Ru-Min Wang$^{1}$\thanks{E-mail: ruminwang@gmail.com},
~Jin-Huan Sheng$^{1}$,~Jie Zhu$^{1}$,~Ying-Ying Fan$^{1}$,~Ya-Dong
Yang$^{2,3}$
  \\
{\scriptsize {$^1$ \it College of Physics and Electronic
Engineering, Xinyang Normal University,
 Xinyang, Henan 464000, China
}}
\\
 {\scriptsize  {$^2$ \it Institute of Particle Physics, Huazhong Normal University, Wuhan,
Hubei 430079, P. R. China }}
\\
 {\scriptsize  {$^3$ \it Key Laboratory of Quark and Lepton Physics,
Ministry of Education, P.R. China }}
 }
 \maketitle
\vspace{-0.4cm}
\begin{abstract}

We study the rare decays $D^+\to \pi^{+}\ell^+\ell^-$, $D^+_s\to
K^{+}\ell^+\ell^-$ and $D^0\to \ell^+\ell^-(\ell=e,\mu)$ in the
minimal supersymmetic standard model with and without R-parity. Using the strong
constraints on relevant supersymmetric parameters from $D^0-\bar{D}^0$ mixing and  $K^+\to \pi^+\nu\bar{\nu}$ decay, we examine constrained supersymmetry
contributions to relevant branching ratios, direct CP violations and ratios of $D^+_{(s)}\to \pi(K)^{+}\mu^+\mu^-$ and $D^+_{(s)}\to \pi(K)^{+}e^+e^-$ decay rates.
We find that both R-parity conserving LR as well as RL mass insertions  and R-parity violating squark exchange couplings have huge effects on the direct
CP violations of $D^+_{(s)}\to \pi(K)^{+}\ell^+\ell^-$, moreover, the constrained  LR and RL mass insertions still have obvious effects on  the ratios of $D^+_{(s)}\to \pi(K)^{+}\mu^+\mu^-$ and $D^+_{(s)}\to \pi(K)^{+}e^+e^-$ decay rates. The direct CP asymmetries and the ratios of $D^+_{(s)}\to \pi(K)^{+}\mu^+\mu^-$ and $D^+_{(s)}\to \pi(K)^{+}e^+e^-$ decay rates are very sensitive to both moduli and
phases of relevant supersymmetric parameters. In addition,  the differential direct CP asymmetries of
$D^+_{(s)}\to \pi(K)^{+}\ell^+\ell^-$  are studied in detail.

\end{abstract}


\newpage

\section{Introduction}
In the standard model (SM), rare decays $D^+\to
\pi^{+}\ell^+\ell^-$, $D^+_s\to K^{+}\ell^+\ell^-$ and $D^0\to
\ell^+\ell^-$ are induced by $c\to u\ell^{+}\ell^{-} $ flavour
changing neutral current (FCNC). Unlike $K$ and $B$ systems, the
short distance (SD) contributions to charmed-meson FCNC processes
are highly suppressed due to the stronger GIM mechanism and weaker
quark mass enhancements in the loops. Therefore, these rare decays
definitely could be good candidates to probe  new physics (NP)
effects. Since the values in the SM are hardly reachable at present
D factories, if any exotic event is found, it must be a strong
evidence for NP.
Many extensions of the SM, such as supersymmetric models
with R-parity violation or models involving a fourth quark
generation, introduce additional diagrams that a priori need not be
suppressed in the same manner as the SM contributions
\cite{Fajfer:2007dy,Buchalla:2008jp,Burdman:2001tf,Burdman:2003rs,Paul:2012ab,Paul:2011br,Jia:2013haa}.

 In the following,  we will concentrate on   $D^+\to
\pi^{+}\ell^+\ell^-$, $D^+_s\to K^{+}\ell^+\ell^-$ and $D^0\to
\ell^+\ell^-$ decays in the minimal supersymmetric standard model
(MSSM) with  and without R-parity.
Using the strong
constraints on relevant supersymmetric parameters from $D^0-\bar{D}^0$ mixing and $K^+\to \pi^+\nu\bar{\nu}$ decay, we will  examine the susceptibilities of relevant
dileptonic invariant mass spectra, branching ratios, differential
direct CP violation, direct CP violation and  ratios of $D^+_{(s)}\to \pi(K)^{+}\mu^+\mu^-$ and $D^+_{(s)}\to \pi(K)^{+}e^+e^-$ decay rates to the constrained supersymmetric coupling parameters.
 Our results indicate that the R-parity violating (RPV) contributions and R-parity conserving (RPC) mass insertion
(MI) contributions could greatly change the direct CP violations of
$D^+_{(s)}\to \pi(K)^{+}\ell^+\ell^-$, furthermore, LR and RL MI contributions also could obviously affect ratios of $D^+_{(s)}\to \pi(K)^{+}\mu^+\mu^-$ and $D^+_{(s)}\to \pi(K)^{+}e^+e^-$ decay rates. These two kinds of  quantities are very sensitive to
both moduli and phases of relevant supersymmetric parameters.  We find that the LD SM effects on
all branching ratios except
$\mathcal{B}(D^0\to \mu^+\mu^-)$ exceed  the largest RPV
contributions, but for the direct CP violations of $D^+_{(s)}\to
\pi(K)^{+}\ell^+\ell^-$, the RPV and RPC contributions could be  totally over the
SM LD ones.

The paper is organized as follows. In Sec. 2, we derive the
expressions for   $D^+_{(s)}\to \pi(K)^{+}\ell^+\ell^-$ and $D^0\to
\ell^+\ell^-$ processes in the MSSM with and without R-parity. In
Sec. 3,  our numerical analysis are presented. We use the constrained parameter spaces from $D^0-\bar{D}^0$ mixing and $K^+\to \pi^+\nu\bar{\nu}$ decay
 to present  the RPV and RPC effects on
 the
observables of the relevant D decays. Sec. 4 is devoted to our
summary.

\section{The theoretical framework for $D^+_{(s)}\to \pi(K)^{+}\ell^+\ell^-$ and $D^0\to \ell^+\ell^-$  decays}
\label{THEORETY}
\subsection{The semileptonic decays $D\to P \ell^+\ell^-$}

 The effective Hamiltonian for the $c\to u\ell^+\ell^-$
transition can be written as
\cite{Lunghi:1999uk,Fajfer:2001sa,Fajfer:2007dy}
\begin{eqnarray}
\mathcal{H}^{SM}_{eff}=-\frac{G_F}{\sqrt{2}}V^{*}_{cb}V_{ub}\sum_{i=7,9,10}(C_{i}\mathcal{O}_i+C'_{i}\mathcal{O'}_i),
\end{eqnarray}
where four-fermion operators are
\begin{eqnarray}
\mathcal{O}_7&=&\frac{e}{8\pi^2}m_cF_{\mu\nu}\bar{u}\sigma^{\mu\nu}(1+\gamma_5)c,\nonumber\\
\mathcal{O}_9&=&\frac{e^2}{16\pi^2}\bar{u}_L\gamma_\mu
c_L\bar{\ell}\gamma^\mu\ell,\nonumber\\
\mathcal{O}_{10}&=&\frac{e^2}{16\pi^2}\bar{u}_L\gamma_\mu
c_L\bar{\ell}\gamma^\mu\gamma_5\ell,\nonumber\\
\mathcal{O}'_7&=&\frac{e}{8\pi^2}m_cF_{\mu\nu}\bar{u}\sigma^{\mu\nu}(1-\gamma_5)c,\nonumber\\
\mathcal{O}'_9&=&\frac{e^2}{16\pi^2}\bar{u}_R\gamma_\mu
c_R\bar{\ell}\gamma^\mu\ell,\nonumber\\
\mathcal{O}'_{10}&=&\frac{e^2}{16\pi^2}\bar{u}_R\gamma_\mu
c_R\bar{\ell}\gamma^\mu\gamma_5\ell,
\end{eqnarray}
with $F_{\mu\nu}$ is the electromagnetic field strength, while
$q_{L(R)}=\frac{1}{2}(1-(+)\gamma_5)q$ is the left(right)-handed
quark field.  The RPC and RPV MSSM effects are including in the
Wilson coefficients corresponding to the $D\to P \ell^+\ell^-$ decay
via
\begin{eqnarray}
C_{i}&=&C^{SM}_i+C^{\spur{R_p}}_{i}+C^{\tilde{g}}_{i},\nonumber\\
C'_{i}&=&C'^{\tilde{g}}_{i},
\end{eqnarray}
and the Wilson coefficients are taken at the scale $\mu=m_c$.

 In the SM,
the expressions of the corresponding Wilson coefficients $C^{SM}_i$
can be found in Refs. \cite{Burdman:2003rs,Greub:1996wn,Isidori:2012yx}.
 In the MSSM
without R-parity, the $c\to u\ell^+\ell^-$ process is mediated by
the tree-level exchange of down squarks
\cite{Fajfer:2007dy,Burdman:2001tf}.
%
The relevant Wilson coefficients
 $C^{\spur{R_p}}_i$ are
\begin{eqnarray}
C^{\spur{R_p}}_{9}=-C^{\spur{R_p}}_{10}=\frac{\sqrt{2}}{G_F}\frac{4\pi}{\alpha_e}\frac{1}{V^{*}_{cb}V_{ub}}\sum_k
\frac{\tilde{\lambda}'_{i2k}\tilde{\lambda}'^{*}_{j1k}}
 {4m^2_{\tilde{d}_{kL}}}.
 \label{RPVH}
\end{eqnarray}
where $\tilde{\lambda}'_{irk}=V^*_{rn}\lambda'_{ink}$
\cite{Petrov:2007gp}, $V^*_{rn}$ is the SM CKM matrix element, and
$\lambda'_{ijk}$ is RPV coupling constant.  Noted that (s)down-down-(s)neutrino vertices have the weak eigenbasis couplings
$\lambda'$, while charged (s)lepton-(s)down-(s)up vertices have the up quark mass eigenbasis
couplings $\tilde{\lambda}'$. Very often in the literature (see e.g. \cite{Golowich:2006gq,Chen:2007dg,Nandi:2006qe,Bhattacharyya:1998be,Kundu:2004cv}), one neglects the difference between
$\lambda'$ and $\tilde{\lambda}'$, based on the fact that diagonal elements of the CKM matrix dominate over nondiagonal
ones.

In the MSSM with R-parity, the gluino-squark exchange coupling give
dominant contributions to these decays. Within the context of the MI
approximation \cite{Gabbiani:1996hi,DFPD-88-TH-8}, allowing for only
one insertion, the relevant Wilson coefficients from the
gluino-squark exchange couplings are \cite{Burdman:2003rs}
\begin{eqnarray}
C^{\tilde{g}}_{7}&=&-\frac{8}{9}\frac{\sqrt{2}}{G_Fm_{\tilde{q}}^2}\pi\alpha_s\left\{(\delta^u_{12})_{LL}\frac{P_{132}(u)}{4}+(\delta^u_{12})_{LR}P_{122}(u)\frac{m_{\tilde{g}}}{m_c}\right\},\nonumber\\
C^{\tilde{g}}_{9}&=&-\frac{8}{27}\frac{\sqrt{2}}{G_Fm_{\tilde{q}}^2}\pi\alpha_s(\delta^u_{12})_{LL}P_{042}(u),\label{RPCWC}
\end{eqnarray}
where $u=m_{\tilde{g}}^2/m_{\tilde{q}}^2$ and the functions
$P_{ijk}(u)$ are defined as
\begin{eqnarray}
P_{ijk}(u)\equiv\int^1_0dx\frac{x^i(1-x)^j}{(1-x+ux)^k}.
\end{eqnarray}
$C'^{\tilde{g}}_{7}$ and $C'^{\tilde{g}}_{9}$ are determined by the
expressions of $C^{\tilde{g}}_{7}$ and $C^{\tilde{g}}_{9}$  with the
replacement $L\leftrightarrow R$, respectively.

 The
decay amplitudes including the SD contribution for $D\to
P\ell^+\ell^-$ decays can be written as follows
\begin{eqnarray}
\mathcal{A}^{SD}(D(p)\to
P(p-q)\ell^+(p_+)\ell^-(p_-))&=&-i\frac{\alpha_eG_F}{4\pi\sqrt{2}}V^*_{cb}V_{ub}f_+(q^2)\left\{(C_{10}+C'_{10})\bar{u}(p_-)\spur{p}\gamma_5v(p_+)\frac{}{}\right.\nonumber\\
&&+\left.\left[(C_{7}+C'_{7})\frac{8m_c}{m_D}+(C_9+C'_9)\right]\bar{u}(p_-)\spur{p}v(p_+)\right\},
\end{eqnarray}
and the form factor $f_+(q^2)$ is defined by  \cite{Isgur:1990kf}
\begin{eqnarray}
\langle
P(p-q)|\bar{u}\gamma^\mu(1\pm\gamma_5)c|D(p)\rangle=(2p-q)^\mu
f_+(q^2)+q^\mu f_-(q^2),~~~~~~~~~~~~~~~~~~~~~~~~~~~~~~~~~~\nonumber\\
\langle
P(p-q)|\bar{u}\sigma^{\mu\nu}(1\pm\gamma_5)c|D(p)\rangle=is(q^2)[(2p-q)^\mu
q^\nu-q^\mu(2p-q)^\nu\pm i\epsilon^{\mu\nu\alpha\beta} (2p-q)_\alpha
q_\beta],
\end{eqnarray}
with the approximation $s(q^2)=f_+(q^2)/m_D$.

Using the formulae presented above,  we  give formulas for the
dilepton invariant mass spectra
\begin{eqnarray}
\frac{d\mathcal{B}^{SD}}{ds}&=&\frac{\tau_D\alpha_e^2G^2_F|V^*_{cb}V_{ub}|^2f^2_+(s)}{2^{14}\pi^5m^3_D}\left\{|C_{10}+C'_{10}|^2
+\left|(C_{7}+C'_{7})\frac{8m_c}{m_D}+(C_{9}+C'_{9})\right|^2\right\}\nonumber\\
&&\left\{\left[(m_D^2-m_P^2+s)^2-4m_D^2s\right]u(s)-\frac{1}{3}u^3(s)\right\},\label{eq:dBds}
\end{eqnarray}
with $s\equiv q^2$.

For the long distance (LD) contributions to the $D^+_{(s)}\to
\pi(K)^{+}\ell^+\ell^-$ decays, we will follow Refs.
\cite{Fajfer:2007dy,Fajfer:2012nr}, the long distance contributions
can be written by the replacements $C_9\rightarrow C_9+C^{LD}$.
\begin{eqnarray}
C^{LD}&=&\frac{i\sqrt{2}}{G_F}\frac{4\pi}{\alpha_e}\frac{1}{V^{*}_{cb}V_{ub}}\big(\frac{a_\rho}{s-m_\rho^2+im_\rho\Gamma_\rho}-\frac{1}{3}\frac{a_\rho}{s-m_\omega^2+im_\omega\Gamma_\omega}\big)\nonumber\\
&&-\frac{32\pi^2}{3}\frac{V^*_{cs}V_{us}}{V^{*}_{cb}V_{ub}}e^{i\delta_\phi}\frac{a_\phi
m_\phi\Gamma_\phi}{s-m_\phi^2+im_\phi\Gamma_\phi}.
\end{eqnarray}
For $D^+\to \pi^+\ell^+\ell^-$, $a_\rho=(2.5\pm0.2)\times10^{-9}$
\cite{Fajfer:2008tm} and  $a_\phi=1.23\pm0.05$ \cite{Fajfer:2012nr}.
For $D^+_s\to K^+\ell^+\ell^-$, $a_\rho=6.97\times10^{-9}$
\cite{Fajfer:2007dy} and $a_\phi=0.49\pm0.05$ \cite{Fajfer:2012nr}.
In addition, $\delta_\phi=0$ will be used in our analyses.

From Eq. (\ref{eq:dBds}), one can obtain the differential direct CP
violation\cite{Fajfer:2012nr,Fajfer:2013bya}
\begin{eqnarray}
a_{CP}(s)&=& \frac{d\mathcal{B}(D^+\to
M^+\ell^+\ell^-)/ds-d\mathcal{B}(D^-\to
M^-\ell^+\ell^-)/ds}{d\mathcal{B}(D^+\to
M^+\ell^+\ell^-)/ds+d\mathcal{B}(D^-\to M^-\ell^+\ell^-)/ds},
\end{eqnarray}
the direct CP violation \cite{Fajfer:2012nr,Fajfer:2013bya}
\begin{eqnarray}
A_{CP}(D^+\to M^+\ell^+\ell^-)&=& \frac{\mathcal{B}(D^+\to
M^+\ell^+\ell^-)-\mathcal{B}(D^-\to
M^-\ell^+\ell^-)}{\mathcal{B}(D^+\to
M^+\ell^+\ell^-)+\mathcal{B}(D^-\to M^-\ell^+\ell^-)},
\end{eqnarray}
and the ratio of decay branching ratios of $D\to P\ell^+\ell^-$ into dimuons over
dielectrons \cite{Hiller:2003js}
\begin{eqnarray}
R_M&=& \frac{\mathcal{B}(D^+\to
M^+\mu^+\mu^-)}{\mathcal{B}(D^+\to
M^+e^+e^-)}.
\end{eqnarray}

\subsection{The leptonic decay $D^0\to \ell^+\ell^-$ }

The general form for the amplitude of $D^0(p)\to
\ell^+(k_+)\ell^-(k_-)$ is \cite{Burdman:2001tf}
\begin{eqnarray}
\mathcal{M}(D^0\to
\ell^+\ell^-)=\bar{u}(k_-)[A_{D^0\ell^+\ell^-}+\gamma_5B_{D^0\ell^+\ell^-}]v(k_+).\label{Eq:pureM}
\end{eqnarray}
The vector leptonic operator $\bar{\ell}\gamma_\mu\ell$  does not
contribute for on-shell leptons as
$p_D^\mu(\bar{\ell}\gamma_\mu\ell)=(p^\mu_{\ell^+}+p^\mu_{\ell^-})(\bar{\ell}\gamma_\mu\ell)=0$,
$i.e.$, $A_{D^0\ell^+\ell^-}=0$. The associated decay branching
ratio is \cite{Golowich:2009ii}
\begin{eqnarray}
\mathcal{B}(D^0\to
\ell^+\ell^-)=\frac{\tau_{D}m_D}{8\pi}\sqrt{1-\frac{4m^2_\ell}{m^2_D}}
\mid B_{D^0\ell^+\ell^-}\mid^2.\label{Eq:pureBr}
\end{eqnarray}

The SM SD contributions in $D^0\to \ell^+\ell^-$ lead to a very
suppressed branching ratio.   $\mathcal{B}^{SD}_{SM}(D^0\to
\mu^+\mu^-)$ is of order $10^{-19}\sim10^{-18}$
\cite{Burdman:2001tf,Paul:2010pq,Pakvasa:1994ni,Gorn:1978sb}, while
taking into the LD contributions, the branching ratio can reach a
level of $10^{-13}$
\cite{Burdman:2001tf,Paul:2010pq,Pakvasa:1994ni}.
$\mathcal{B}^{SD}_{SM}(D^0\to e^+e^-)$ is of order $10^{-23}$
\cite{Burdman:2001tf}, which is much smaller than
$\mathcal{B}^{SD}_{SM}(D^0\to \mu^+\mu^-)$. These smallness of the
SM signal makes it easier for NP contributions to stand out.

In the RPV MSSM, the coefficient $B^{\spur{R_p}}_{D^0\ell^+\ell^-}$
given in Eq. (\ref{Eq:pureM}) is \cite{Golowich:2009ii}
\begin{eqnarray}
 B^{\spur{R_p}}_{D^0\ell^+\ell^-}&=&\sum_k
\frac{\tilde{\lambda}'_{i2k}\tilde{\lambda}'^{*}_{j1k}}
 {4m^2_{\tilde{d}_{kL}}}im_\ell f_D.
\end{eqnarray}
In the RPC MSSM, the squark-gluino contribution to the pure leptonic
D decays \cite{Burdman:2001tf,Lunghi:1999uk}
\begin{eqnarray}
 B^{\tilde{g}}_{D^0\ell^+\ell^-}&=&\frac{i\sqrt{2}G_F}{9\pi}\alpha_s m_\ell f_D
 \left[(\delta^u_{22})_{LR}(\delta^u_{12})_{RL}P_{032}-(\delta^u_{22})_{RL}(\delta^u_{12})_{LR}P_{122}\right].
\end{eqnarray}
The double MI is required to induce a helicity flip in the squark
propagator. Noted that the chargino contribution to the $Z$ penguin
for $D\to \ell^+ \ell^-$ also contains a double MI. Due to the
double MI, these contributions to $D\to \ell^+ \ell^-$ is completely
negligible, and we will not examine the MI effects in $D\to \ell^+
\ell^-$ decays.

\section{Numerical results and analyses}
With the formulae  presented in previous section, we are ready to
perform our numerical analysis.
 When we study the effects due to the MSSM with and without R-parity,
 we consider only one new coupling at one time, neglecting the interferences between
different new couplings, but keeping their interferences with the SM
amplitude. The input parameters are collected in the Appendix.  The
experimental upper limits at 90\% confidence level (CL)
\cite{Aaij:2013sua,Aaij:2013cza,PDG2014} are listed in the second
column of Tab. \ref{Tab:EXPpredictions}.
and the SM predictions excluding
(including) LD contributions are also listed in the third (last)
column of  Tab. \ref{Tab:EXPpredictions}. The input parameters varied
randomly within $1\sigma$  variance will be used in this work.

\begin{table}[th]
\caption{The experimentalupper limits
\cite{Aaij:2013sua,Aaij:2013cza,Lees:2011hb,Petric:2010yt,PDG2014}
and the theoretical predictions with 1$\sigma$ error ranges of the
input parameters for relevant branching ratios (in units of $10^{-10}$).}\label{Tab:EXPpredictions}
\vspace{-0.5cm}
\begin{center}{\footnotesize
\begin{tabular}{l|l|l|l|l|l|l}\hline\hline
 Modes&~Best exp.&\multicolumn{4}{c}{SD contributions}\vline& LD contributions \\ \cline{3-6}
&~limits&~~~~~SM  &~~~~ S1\footnotemark[1]&~~~~~~S2\footnotemark[1]&~~~~~~S3\footnotemark[1]& ~~~~~~~~~SM\\
\hline
$\mathcal{B}(D_u^0\to
e^+e^-)$&$<790$& $\approx10^{-13}$ & $\leq2.65\times10^{-7}$ &$~~~~~\cdots$&$~~~~~\cdots$&$[0.0027,0.008]$\\
$\mathcal{B}(D_d^+\to \pi^+e^+e^-)$&$<11000$&$[4.84,8.46]$&$[4.03,9.31]$&$[1.10,19.68]$&$[1.02,21.06]$&$[920,1200]$\\
$\mathcal{B}(D_s^+\to K^+e^+e^-)$&$<37000$&$[1.77,3.08]$&$[1.38,3.28]$&$[1.00,7.42]$&$[1.00,7.66]$&$[2090,2220]$\\
\hline
$\mathcal{B}(D_u^0\to \mu^+\mu^-)$&$<62$&$10^{-9}\sim10^{-8}$&$\leq0.011$&$~~~~~\cdots$&$~~~~~\cdots$&$[0.0027,0.008]$\\
$\mathcal{B}(D_d^+\to \pi^+\mu^+\mu^-)$&$<730$&$[4.59,8.04]$&$[3.80,8.80]$&$[1.06,18.76]$&$[1.03,20.10]$&$[920,1200]$\\
$\mathcal{B}(D_d^+\to \pi^+\mu^+\mu^-)_{l}\footnotemark[2]$&$<200$&$[0.72,1.27]$&$[0.60,1.40]$&$[0.10,3.00]$&$[0.10,3.22]$&$[0.72,1.27]$\\
$\mathcal{B}(D_d^+\to \pi^+\mu^+\mu^-)_{h}\footnotemark[2]$&$<260$&$[1.18,2.08]$&$[0.99,2.28]$&$[0.18,4.81]$&$[0.17,5.15]$&$[1.18,2.08]$\\
$\mathcal{B}(D_s^+\to K^+\mu^+\mu^-)$&$<210000$&$[1.64,2.84]$&$[1.26,3.02]$&$[1.00,6.93]$&$[1.00,7.17]$&$[2090,2220]$\\
\hline \hline
\end{tabular}}
\end{center}
\end{table}
\footnotetext[1]{S1 denotes the RPV predictions constrained from $K^+\to \pi^+\nu\bar{\nu}$ decay and $D^0-\bar{D}^0$ mixing at 99.7\% CL, S2 and S3 denote the RPC predictions constrained from $D^0-\bar{D}^0$ mixing at 90\% and 99.7\% CL, respectively.}
\footnotetext[2]{
$\mathcal{B}(D_d^+\to \pi^+\mu^+\mu^-)_{l}$ denotes the branching
ratios in $s\in[0.250,0.525]$ GeV$^2$ bin, and $\mathcal{B}(D_d^+\to
\pi^+\mu^+\mu^-)_{h}$ denotes  the branching ratios in
$s\in[1.250,2.000]$ GeV$^2$ bin.}

\subsection{RPV MSSM effects}

First, we will  consider the RPV effects and  further constrain the
relevant RPV couplings from  relevant experimental data. As given in Sec.
\ref{THEORETY}, there is only one RPV coupling product
$\tilde{\lambda}'_{12k}\tilde{\lambda}'^{*}_{11k}$
($\tilde{\lambda}'_{22k}\tilde{\lambda}'^{*}_{21k}$)  relevant to
$c\to u e^+e^-$ ($c\to u \mu^+\mu^-$) transition.
 Noted that $\tilde{\lambda}'_{i2k}\tilde{\lambda}'^{*}_{i1k}$ and $\lambda'_{i2k}\lambda'^{*}_{i1k}$ also contribute to $D^0-\bar{D}^0$ mixing and $K^+\to \pi^+\nu\bar{\nu}$ decay, respectively. In this work, we neglect the difference between
$\lambda'$ and $\tilde{\lambda}'$, and the bounds of relevant RPV coupling products from $D^0-\bar{D}^0$ mixing and $K^+\to \pi^+\nu\bar{\nu}$ decay are considered.

\begin{figure}[t]
\begin{center}
\includegraphics[scale=1.6]{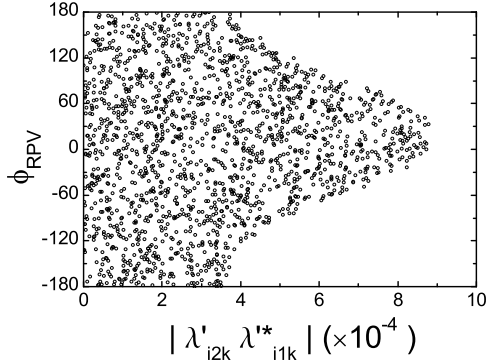}
\end{center}
\vspace{-0.4cm}
 \caption{ The allowed RPV parameter spaces  from $K^+\to\pi^+\nu\bar{\nu}$ at 99.7\% CL with 1000 GeV sfermions,
 and the RPV weak phase $(\phi_{RPV})$ is given in degree.}
 \label{fig:bounds}
\end{figure}

Relevant expressions of $D^0-\bar{D}^0$ mixing  and $K^+\to \pi^+\nu\bar{\nu}$ decay can be found in Ref. \cite{Golowich:2007ka} and Ref. \cite{Marciano:1996wy,Deandrea:2004ae,Buras:2004uu}, respectively.
The latest $D^0-\bar{D}^0$ mixing parameters  $x_D=(0.56\pm0.19)\%$ \cite{Ko:2014jda}  and $\mathcal{B}(K^+\to \pi^+\nu\bar{\nu})=(1.7\pm1.1)\times10^{-10}$ \cite{PDG2014}  will be used to constrain the RPV coupling products $\tilde{\lambda}'_{i2k}\tilde{\lambda}'^{*}_{i1k}$. Assuming the RPV
coupling products
are complex, $m_{\tilde{q}}=1000$ GeV and $m^2_{\tilde{g}}/m^2_{\tilde{q}}\in[0.25,4]$, we obtained $|\tilde{\lambda}'_{i2k}\tilde{\lambda}'^{*}_{i1k}|\in[0.0058,0.019]$ from $D^0-\bar{D}^0$ mixing  and $|\lambda'_{i2k}\lambda'^{*}_{i1k}|\leq7.95\times10^{-4}$ from $K^+\to \pi^+\nu\bar{\nu}$ decay within 1.64 experimental standard deviations (at 90\% CL), in other words, the RPV coupling products are excluded at 90\% CL. Within three standard deviations (at 99.7\% CL), we get $|\tilde{\lambda}'_{i2k}\tilde{\lambda}'^{*}_{i1k}|\leq0.023$ and corresponding weak phase is free from $D^0-\bar{D}^0$ mixing,  furthermore, $|\lambda'_{i2k}\lambda'^{*}_{i1k}|\leq8.97\times10^{-4}$ and the phase is also constrained  from $K^+\to \pi^+\nu\bar{\nu}$ decay.  Considering the experimental bounds of $D^0-\bar{D}^0$ mixing and $K^+\to \pi^+\nu\bar{\nu}$ decay at 99.7\% CL at the same time, the effective constraints from $K^+\to \pi^+\nu\bar{\nu}$ decay at 99.7\% CL, which is shown in Fig. \ref{fig:bounds} and  will be used to study the RPV effects in
$D^+_{(s)}\to \pi(K)^{+}\ell^+\ell^-$ and $D^0\to \ell^+\ell^-$ decays. Noted that the experimental upper limits at 90\% CL listed in the second
column of Tab. \ref{Tab:EXPpredictions} do not give any further constraints on relevant RPV couplings.

Now we will use  the constrained RPV coupling space  from $K^+\to \pi^+\nu\bar{\nu}$ decay at 99.7\% CL
shown in Fig. \ref{fig:bounds} to explore the RPV coupling effects
in $D_u^0\to \ell^+\ell^-$, $D_d^+\to \pi^+\ell^+\ell^-$ and
$D_s^+\to K^+\ell^+\ell^-$ decays.
RPV numerical results of the branching ratios  are summarized in the forth column of Tab.
\ref{Tab:EXPpredictions}.  Comparing to the SM predictions excluding LD contributions , one can find
that the constrained RPV couplings still have great effects on $\mathcal{B}(D_u^0\to \ell^+\ell^-)$, nevertheless, other four semileptionic branching ratios are not very obviously affected by the constrained RPV couplings.
Comparing to the SM predictions including LD contributions in the
last column of Tab. \ref{Tab:EXPpredictions},  the constrained
RPV contributions may a little  larger than the LD ones to $\mathcal{B}(D_u^0\to \mu^+\mu^-)$.
Comparing to the experimental upper
limits given in the second column of Tab. \ref{Tab:EXPpredictions},
one can find that all experimental upper limits are much larger than the theoretical prediction of branching ratios with SD contributions, so all upper limits listed in  the second column of Tab. \ref{Tab:EXPpredictions} do not give any further constraint to relevant RPV coupling products.

\begin{figure}[t]
\begin{center}
\includegraphics[scale=1.3]{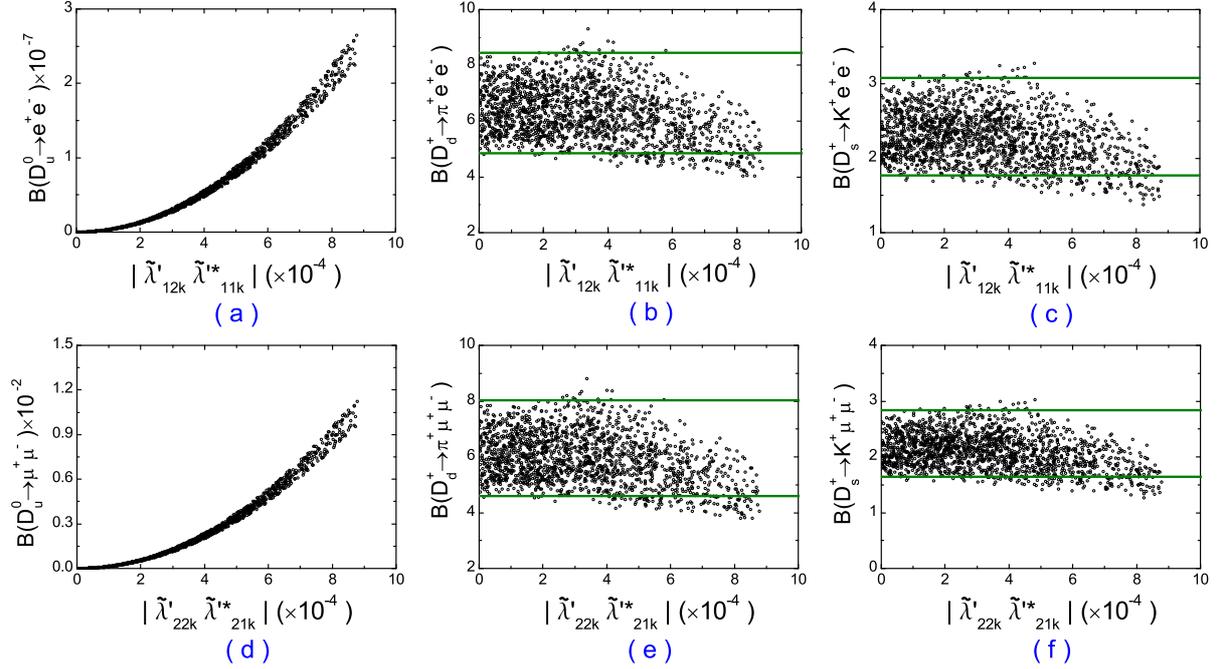}
\end{center}
\vspace{-0.4cm}
 \caption{ The RPV coupling effects  due to the squark exchange
 on the branching ratios (in units of $10^{-10}$), and  the green horizontal solid lines
represent the SM ranges with 1¦Ò error of the input parameters (the same in Fig. \ref{fig:ACPe},  Fig. \ref{fig:ACPmu} and Fig. \ref{fig:brACPLR}).}
 \label{fig:br}
\end{figure}

The sensitivities of the branching ratios to the moduli of the
constrained RPV coupling products are shown in  Fig. \ref{fig:br}.
 Fig. \ref{fig:br} show us that $\mathcal{B}(D_u^0\to \ell^+\ell^-)$  are very sensitive to the moduli of squark exchange
RPV couplings, and they are great increasing with
$|\tilde{\lambda}'_{12k}\tilde{\lambda}'^{*}_{11k}|$ or
$|\tilde{\lambda}'_{22k}\tilde{\lambda}'^{*}_{21k}|$.
Other four branching ratios of semileptonic decays are not very sensitive to the moduli of squark exchange
RPV couplings. In addition, all branching ratios
are not sensitive to the relevant RPV weak phases.

\begin{figure}[ht]
\begin{center}
\includegraphics[scale=1.3]{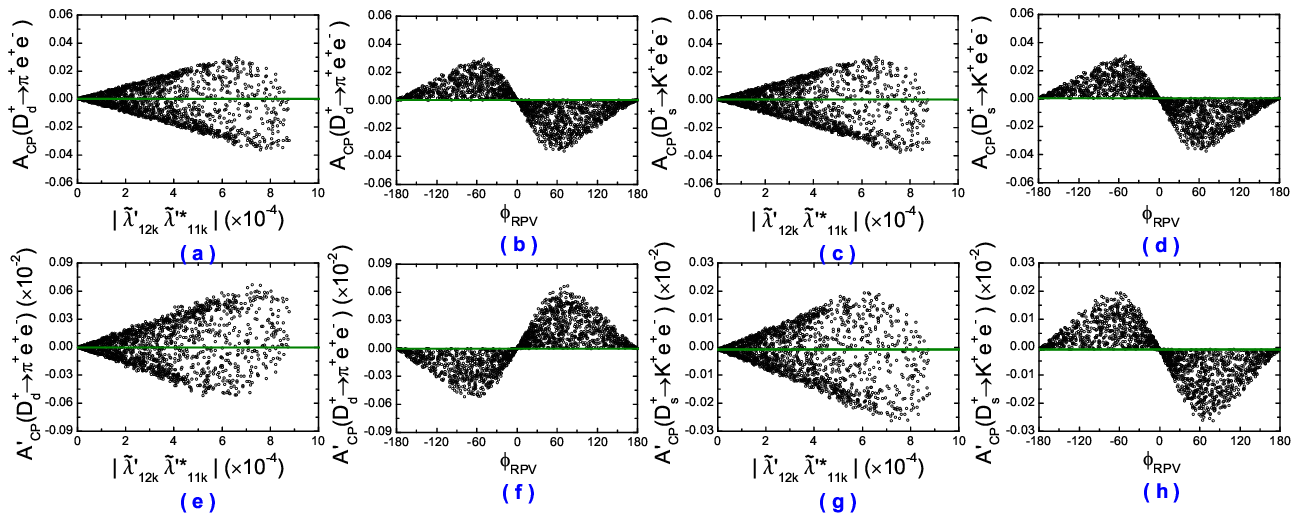}
\end{center}
\vspace{-0.4cm}
 \caption{ The effects of RPV coupling $\tilde{\lambda}'_{12k}\tilde{\lambda}'^*_{11k}$ due to the squark exchange
 on the direct CP violations of
 $D^+_d\to \pi^+e^+e^-$ and $D^+_s\to K^+e^+e^-$ decays. }
 \label{fig:ACPe}
\begin{center}
\includegraphics[scale=1.3]{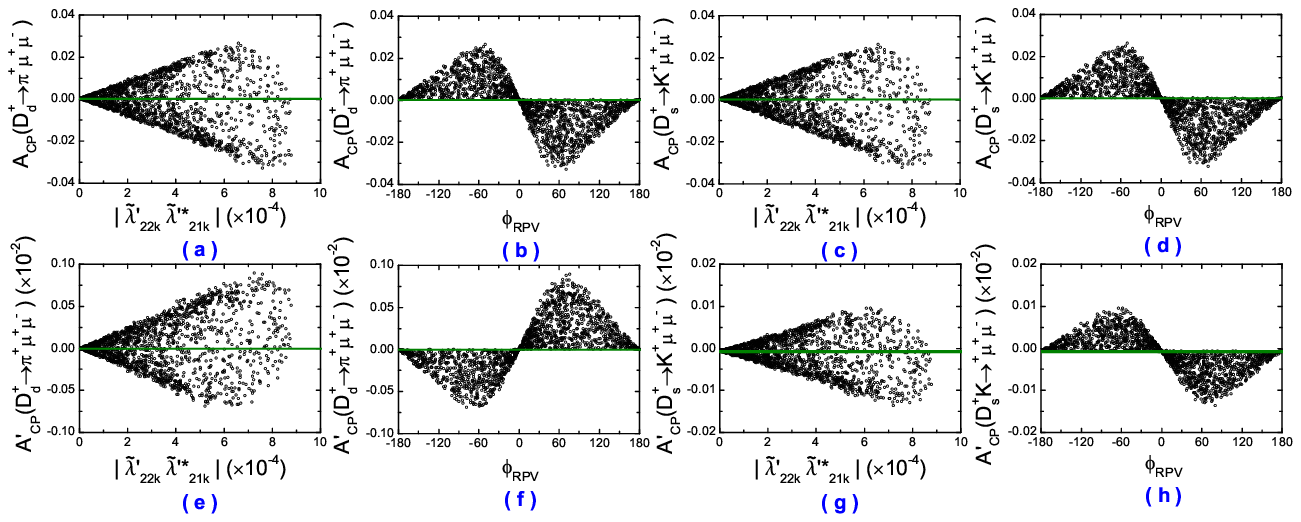}
\end{center}
\vspace{-0.4cm}
 \caption{ The effects of RPV coupling $\tilde{\lambda}'_{22k}\tilde{\lambda}'^*_{21k}$ due to the squark exchange
on the direct CP violations of
 $D^+_d\to \pi^+\mu^+\mu^-$ and $D^+_s\to K^+\mu^+\mu^-$ decays. }
 \label{fig:ACPmu}
\end{figure}

Our main purpose is studying the  direct CP violations of semileptonic D decays. Many theoretical uncertainties such as involved in the form factors, decay constants and CKM matrix elements are cancelled  in the  ratios.
The RPV predictions of the direct CP violations are very dependent on the RPV coupling products, so we will not give the numerical results and only show their sensitivities to the
constrained RPV coupling products.
The correlations between relevant direct CP violations and the
constrained RPV coupling products are shown in Fig. \ref{fig:ACPe}
and Fig. \ref{fig:ACPmu} for  $D^+_{(s)}\to \pi(K)^+e^+e^-$ and
$D^+_{(s)}\to \pi(K)^+\mu^+\mu^-$, respectively,
the direct CP violation excluding (including) the LD contributions
are denoted by $A_{CP}~ (A'_{CP})$, and the SM ranges of relevant
direct CP violations are displayed by green horizontal solid lines.
 In the SM,  the direct CP violations
of these four semileptonic D decays are tiny, and they are at about
$10^{-4} ~(10^{-6})$ order if LD contributions excluded (included).
As shown in Fig. \ref{fig:ACPe} and Fig. \ref{fig:ACPmu}, the
constrained RPV couplings still have huge effects on the direct CP
violations in the case of both excluded and included LD
contributions, all direct CP violations could be enhanced about two orders of magnitude, and they
are very sensitive to
both moduli and weak phases of the RPV coupling products. In
addition, the interference of resonant part of the LD contribution
and the NP affected SD contribution could reduce the direct CP
violations more than  one order of magnitude.

\begin{figure}[t]
\begin{center}
\includegraphics[scale=1.3]{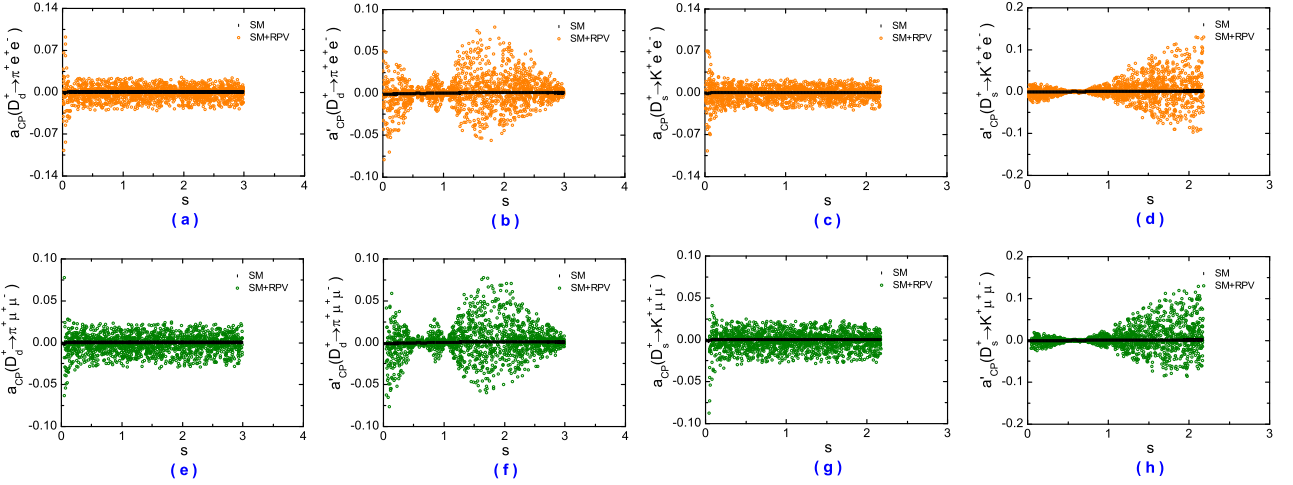}
\end{center}
\vspace{-0.4cm}
 \caption{ The RPV coupling  effects  due to the squark exchange
 in  the differential direct CP violations  of
 $D^+_d\to \pi^+\ell^+\ell^-$ and $D^+_s\to K^+\ell^+\ell^-$ decays.}
 \label{fig:aCP}
\end{figure}

Furthermore, the constrained RPV coupling effects on  the differential direct CP violations of
$D^+_{(s)}\to \pi(K)^{+}\ell^+\ell^-$ decays are also displayed in
Fig. \ref{fig:aCP}, and the SM predictions are given for comparison.
The differential direct
CP violations excluding (including) the LD contributions
are denoted by $a_{CP}$ ($a'_{CP}$).
Fig. \ref{fig:aCP} shows us that the RPV couplings could have huge
contributions to $a_{CP}(D^+_{(s)}\to \pi(K)^+\ell^+\ell^-)$
at almost all $s$ regions, and at middle  $s$ regions, the interference of resonant part of the LD contribution
and the NP affected SD contribution could obviously reduce $a'_{CP}(D^+_{(s)}\to \pi(K)^+\ell^+\ell^-)$.
 At all $s$ regions, we get that $a^{SM}_{CP}(D^+_{d}\to
 \pi^+e^+e^-)\in[-1.7\times10^{-3},4.67\times10^{-4}]$ and  $a^{SM}_{CP}(D^+_{s}\to
K^+e^+e^-)\in[-1.6\times10^{-3},4.76\times10^{-4}]$. The constrained
RPV couplings could enhance $a^{(')}_{CP}$ about two orders of
magnitude from their SM predictions.
Noted that,  for the strong phase $\delta_\phi$ on the $\phi$
resonant peak appeared in the LD amplitude,
$\delta_\phi=\frac{\pi}{2},\pi$ cases also have been studied in Ref.
\cite{Fajfer:2012nr}, $a'_{CP}$ for $\delta_\phi=\frac{\pi}{2}$
maybe larger a little than one for $\delta_\phi=0,\frac{\pi}{2}$.
Nevertheless, the enhancements of the RPV contributions are still
totally over the LD contributions in three $\delta_\phi$ cases.

\subsection{RPC MSSM effects}
Now we turn to the gluino-mediated MSSM effects in $D^+_{(s)}\to
\pi(K)^{+}\ell^+\ell^-$ decays in the framework of the MI
approximation.
 The effects of the constrained LL
insertion on $D^+_{(s)}\to \pi(K)^{+}\ell^+\ell^-$ decays are almost
negligible because of lacking the gluino mass enhancement in the
decay,  and they will not provide any significant effect on the
branching ratios and the direct CP violations of  $D^+_{(s)}\to
\pi(K)^{+}\ell^+\ell^-$ decays.
The LR and RL MIs only generate magnetic operators
$\mathcal{O}_{7\gamma}$ and $\mathcal{O}'_{7\gamma}$, respectively.
Since LR and RL insertion contributions are enhanced by
$m_{\tilde{g}}/m_b$ due to the chirality flip from the gluino in the
loop compared with the contribution including the SM one,  even a
small $(\delta_{12}^u)_{LR}$ or $(\delta_{12}^u)_{RL}$ may  have
large effects in the decay.  So we will only consider the LR and RL MI effects in this work.

Since the most stringent bounds come from
$D^0-\bar{D}^0$ mixing, we will take into account the constraints
set by the $D^0-\bar{D}^0$ mixing to
investigate NP contributions in  $D^+_{(s)}\to
\pi(K)^{+}\ell^+\ell^-$ decays.
The latest $D^0-\bar{D}^0$ mixing parameters, $x_D=(0.56\pm0.19)\%$ \cite{Ko:2014jda}, will be used to constrain the LR and RL MIs. Using the formulae in Ref. \cite{Golowich:2007ka}, we get
$|(\delta^u_{12})_{LR,RL}|\in[0.010,0.034]$ at 90\% CL and $|(\delta^u_{12})_{LR,RL}|\leq0.037$ at 99.7\% CL  with
$m_{\tilde{q}}=1000$ GeV and $m^2_{\tilde{g}}/m^2_{\tilde{q}}\in[0.25,4]$.
Nevertheless, the phases of  $(\delta^u_{12})_{LR,RL}$ are not restricted.

Because the LR and RL MI effects in
$D^+_{(s)}\to \pi(K)^{+}e^+e^-$ and $D^+_{(s)}\to
\pi(K)^{+}\mu^+\mu^-$ are same as each other, here we take
the LR insertion effects in $D^+_{(s)}\to \pi(K)^{+}\mu^+\mu^-$ as
an example.
RPC numerical results of the branching ratios constrained from $D^0-\bar{D}^0$ mixing at 90\% and 99.7\% CL  are summarized in the fifth and sixth columns of Tab.
\ref{Tab:EXPpredictions}, respectively.  The constrained RPC MI couplings still have obvious effects on four semileptonic branching ratios, but their contributions  are much smaller than the LD ones.

\begin{figure}[t]
\begin{center}
\includegraphics[scale=1.3]{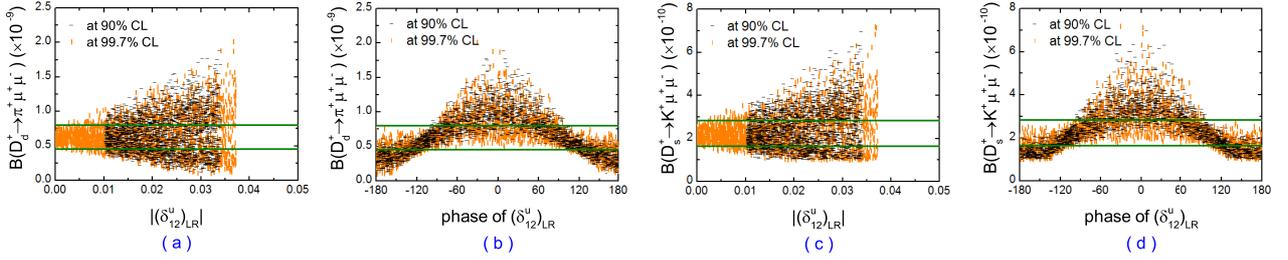}
\end{center}
\vspace{-0.4cm}
 \caption{The LR MI effects
 on the branching ratios of
 $D^+_d\to \pi^+\mu^+\mu^-$ and $D^+_s\to K^+\mu^+\mu^-$ decays.}
 \label{fig:brLR}
\end{figure}

\begin{figure}[htb]
\begin{center}
\includegraphics[scale=1.3]{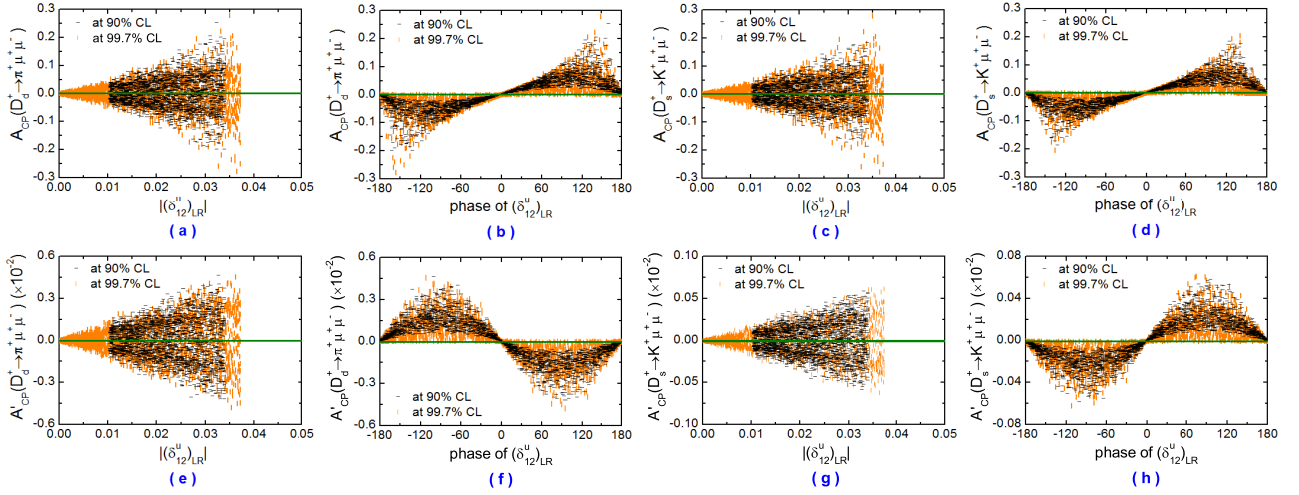}
\end{center}
\vspace{-0.4cm}
 \caption{The LR MI effects
 on  the direct CP violations of
 $D^+_d\to \pi^+\mu^+\mu^-$ and $D^+_s\to K^+\mu^+\mu^-$ decays.}
 \label{fig:ACPLR}
\end{figure}

The LR MI effects on $\mathcal{B}(D^+_{(s)}\to
\pi(K)^{+}\mu^+\mu^-)$ and $A_{CP}(D^+_{(s)}\to
\pi(K)^{+}\mu^+\mu^-)$ are shown in Fig. \ref{fig:brLR} and Fig. \ref{fig:ACPLR}, respectively.
As shown in Fig. \ref{fig:ACPLR},  $A_{CP}(D^+_{(s)}\to
\pi(K)^{+}\mu^+\mu^-)$ are obviously affected by the constrained LR
MI constrained from  $D^0-\bar{D}^0$ mixing, their RPC predictions could be much larger than their SM ones. In addition, the branching ratios and the direct CP violations  are quite sensitive to
 both modulus and weak phase of $(\delta_{12}^u)_{LR}$.

\begin{figure}[t]
\begin{center}
\includegraphics[scale=1.3]{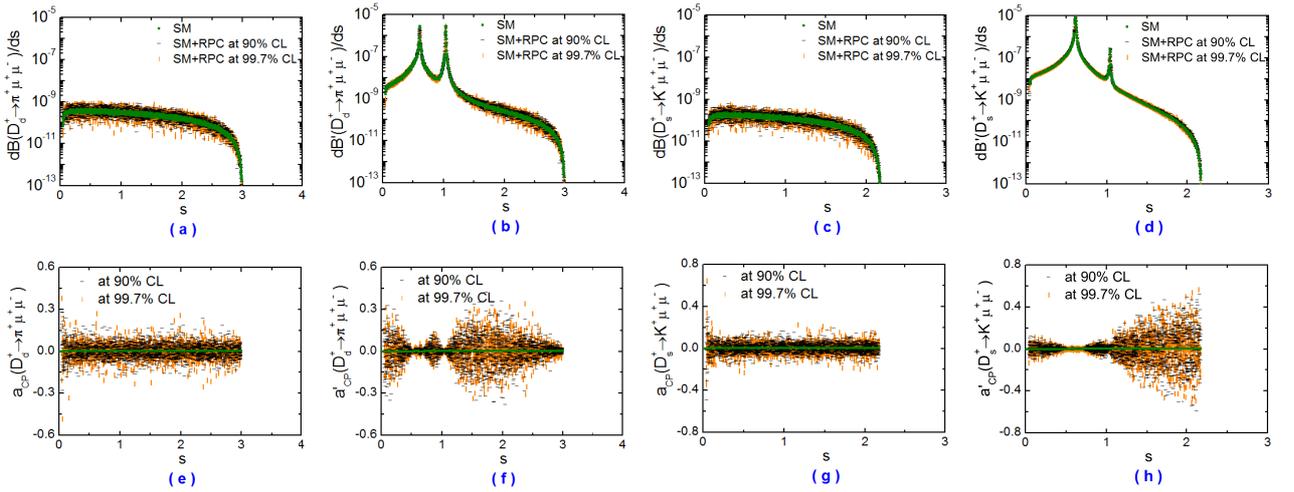}
\end{center}
\vspace{-0.4cm}
 \caption{The LR MI effects on the differential branching ratios and the differential direct CP violations of
 $D^+_d\to \pi^+\mu^+\mu^-$ and $D^+_s\to K^+\mu^+\mu^-$ decays.}
 \label{fig:dbrdacpLR}
\end{figure}

The constrained LR MI effects on  the differential branching ratios and the differential direct CP violations  of  $D^+_d\to \pi^+\mu^+\mu^-$ and
$D^+_s\to K^+\mu^+\mu^-$ decays are displayed in Fig.
\ref{fig:dbrdacpLR}.
As shown in Fig. \ref{fig:dbrdacpLR} (a-d), the differential branching
ratios could be slightly
affected by the constrained LR MI. From Fig. \ref{fig:dbrdacpLR} (e) and (g), one can see that  the differential direct CP violations
could be significantly affected at all $s$ regions by the constrained LR MI, nevertheless, as displayed at middle $s$ region of Fig. \ref{fig:dbrdacpLR} (f) and (h), the interference of resonant part of
the LD contribution and the RPC affected SD contribution could
hugely reduce the differential direct CP violations more than  two
orders of magnitude.

\begin{figure}[t]
\begin{center}
\includegraphics[scale=1]{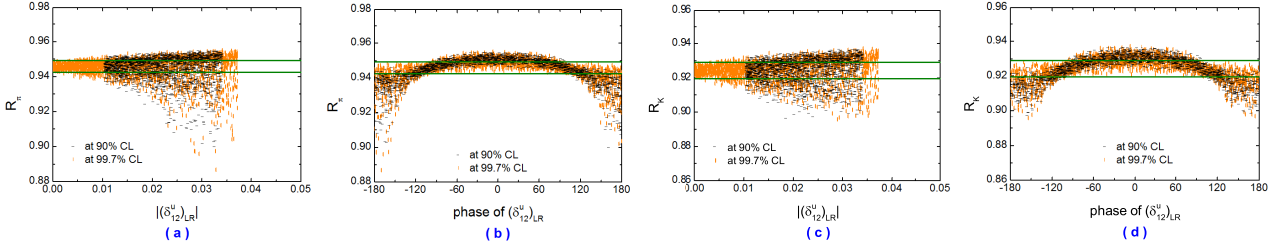}
\end{center}
\vspace{-0.4cm}
 \caption{The LR MI effects on the ratios of  $\mathcal{B}(D^+_{(s)}\to\pi(K)^{+}\mu^+\mu^-)$ and $\mathcal{B}(D^+_{(s)}\to\pi(K)^{+}e^+e^-)$  decays.}
 \label{fig:R}
\end{figure}

Moreover, the ratios of decay branching ratios of $D^+_{(s)}\to\pi(K)^{+}\ell^+\ell^-$ into dimuons over
dielectrons  are also studied since the branching ratios of $D^+_{(s)}\to\pi(K)^{+}\ell^+\ell^-$ are affected by the same LR and RL MI parameters. The constrained LR MI effects on  $R_\pi$ and $R_K$ are displayed in Fig. \ref{fig:R}. One can see that the constrained LR MI still has great effects on  these ratios, which are very sensitive to
 both modulus and weak phase of $(\delta_{12}^u)_{LR}$. When the modulus and absolute phase of $(\delta_{12}^u)_{LR}$ are large, $R_\pi$ and $R_K$ may have small values.

\section{Summary}

In this work we have performed a brief study of the RPV coupling
effects and the RPC MI effects in the MSSM from $D^+_{(s)}\to
\pi(K)^{+}e^+e^-,\pi(K)^{+}\mu^+\mu^-$ and $D^0\to
e^+e^-,\mu^+\mu^-$ decays. Considering the theoretical
uncertainties and using the strong
constraints on relevant supersymmetric parameters from $D^0-\bar{D}^0$ mixing or $K^+\to \pi^+\nu\bar{\nu}$ decay, we have investigated the
sensitivities of the dileptonic  invariant mass spectra, branching
ratios, differential direct CP violation, direct CP violation and ratios of $D^+_{(s)}\to \pi(K)^{+}\mu^+\mu^-$ and $D^+_{(s)}\to \pi(K)^{+}e^+e^-$ decay rates to
the survived RPV and RPC coupling spaces.

We have found that, after satisfying the current experimental
bounds of $D^0-\bar{D}^0$ mixing and $K^+\to \pi^+\nu\bar{\nu}$ decay at 99.7\% CL,  left-handed squark exchange RPV
couplings have significant effects on $\mathcal{B}(D^0\to
\ell^+\ell^-)$, differential direct CP violation and
direct CP violation of $D^+_{(s)}\to
\pi(K)^{+}\ell^+\ell^-$ semileptonic decays.
The direct CP violations of
$D^+_{(s)}\to \pi(K)^{+}\ell^+\ell^-$ are sensitive to both moduli
and phases of relevant RPV coupling products.

As for the RPC MI effects, we found that the constrained LR and RL
insertions from $D^0-\bar{D}^0$ mixing at 90\% CL and 99.7\% CL  could significantly affect
the branching ratios and the direct CP violations of  $D^+_{(s)}\to
\pi(K)^{+}\ell^+\ell^-$  as well as ratios of $D^+_{(s)}\to \pi(K)^{+}\mu^+\mu^-$ and $D^+_{(s)}\to \pi(K)^{+}e^+e^-$ decay rates, and they are very sensitive to both moduli and phases
of relevant LR and RL mass insertion couplings.

 With the running  of LHC experiment, the prospects of measuring $D^+_{(s)}\to \pi(K)^{+}\ell^+\ell^-$ and
$D^0\to \ell^+\ell^-$  decays   could be  realistic.  We expect that
future experiments will significantly strengthen the allowed
parameter spaces for RPV couplings and RPC MIs. Our predictions  on
related the direct CP violations of semileptonic D decays and the  ratios of $D^+_{(s)}\to \pi(K)^{+}\mu^+\mu^-$ and $D^+_{(s)}\to \pi(K)^{+}e^+e^-$ decay rates  could be very useful for probing supersymmetric
effects  in future experiments.

\section*{Acknowledgments}

The work was supported by  Program for the New
Century Excellent Talents in University (No. NCET-12-0698) and the National Natural Science Foundation of
China (Nos. 11105115 and 1122552).

\begin{appendix}
\section*{Appendix: Input parameters}
\label{SEC.INPUT}

The input parameters are collected in Table \ref{Tab.input}. We have
several remarks on the input parameters:
\begin{itemize}
\item \underline{Wilson coefficients}: The SM Wilson coefficients $C^{SM}_i$ are obtained from the
expressions in Ref. \cite{Burdman:2003rs}.

\item \underline{CKM matrix element}:  For the SM predictions,
we use the CKM matrix elements from the Wolfenstein parameters of
the latest analysis within the SM in Ref. \cite{UTfit}, and   for
the SUSY predictions, we take the CKM matrix elements in terms of
the Wolfenstein parameters of the NP generalized analysis results in
Ref. \cite{UTfit}.

\item  \underline{Form factors}:  %
We  use the results of form factors in Ref. \cite{Melikhov:2000yu}.
In our numerical data analysis, the 10\% uncertainties induced by
$F(0)$ are also considered.

 \end{itemize}

\begin{table}[h]
\caption{Values of the theoretical input parameters. To be
conservative, we use all theoretical input parameters at 68\% CL in
our numerical results.} {\footnotesize
\begin{center}
\begin{tabular}{lr}\hline\hline
$m_W=80.385\pm 0.015~{\rm GeV},~m_{D_u}=1.865~{\rm GeV},~m_{_{D_d}}=1.870~{\rm GeV},~m_{_{D_s}}=1.969~{\rm GeV},$\\
$m_{D^*_d}=0.2.010~{\rm GeV},~m_{D^*_s}=2.112~{\rm GeV},~m_{\pi^+}=0.140~{\rm GeV},~m_{K^{+}}=0.494~{\rm GeV},~m_{\rho^0}=0.775~{\rm GeV},$\\
$m_{\omega}=0.783~{\rm GeV},~m_{\phi}=1.019~{\rm
GeV},~m_t=173.5\pm1.0~{\rm GeV},~m_b=4.78\pm0.06~{\rm
GeV},$\\
 $m_c=1.67\pm0.07~{\rm
GeV},~\overline{m}_b(\overline{m}_b)=(4.18\pm0.03)~{\rm
GeV},~\overline{m}_s(2{\rm GeV})=(0.095\pm0.005)~{\rm
GeV},$\\
$\overline{m}_d(2{\rm GeV})=(0.0048\pm0.0003)~{\rm
GeV},~m_e=5.11\times10^{-4}~{\rm GeV},~m_\mu=0.106~{\rm GeV};$\\
$\tau_{D_u}=(0.4101\pm0.0015)~{\rm
ps},~\tau_{D_d}=(1.040\pm0.007)~{\rm
ps},~\tau_{D_s}=(0.500\pm0.007)~{\rm ps},$\\
$\Gamma_{\rho^0}=(0.1462\pm0.0007)~{\rm
GeV},~\Gamma_{\omega}=(8.49\pm0.08)\times10^{-3}~{\rm
GeV},~\Gamma_{\phi}=(4.26\pm0.04)\times10^{-3}~{\rm GeV};$\\
$f_{D_u}=0.2067\pm0.0089~{\rm GeV},~f_{K^+}=(0.15610\pm0.00083)~{\rm
GeV},~f_{\pi}=(0.13041\pm0.00020)~{\rm GeV}.$
&\cite{PDG2014}\\\hline
The Wolfenstein parameters for the SM predictions: &\\
$A=0.827\pm0.013,~\lambda=0.22535\pm0.00065,~\bar{\rho}=0.132\pm0.021,~\bar{\eta}=0.350\pm0.014$.&\\
The Wolfenstein parameters for the SUSY predictions: & \\
$A=0.802\pm0.020,~\lambda=0.22535\pm0.00065,~\bar{\rho}=0.147\pm0.048,~\bar{\eta}=0.370\pm0.057$.&\cite{UTfit}\\
\hline

\end{tabular}
\end{center}}\label{Tab.input}
\end{table}

\end{appendix}

\end{document}